\documentclass[final,3p]{elsarticle}

\usepackage{graphicx}

\usepackage{bm,amsmath,multirow}

\usepackage{amssymb}

\journal{Physica C}

\begin{document}

\begin{frontmatter}


 \author{Ai Yamakage \corref{cor1}}
 \ead{ai@rover.nuap.nagoya-u.ac.jp}
\cortext[cor1]{Corresponding author. Address: Department of Applied Physics, Nagoya University, Nagoya 464-8603, Japan. Tel.: +81 52 798 3701; fax: +81 52 789 3298.}

\title{Theory of tunneling spectroscopy in a superconducting topological insulator}


\author{Keiji Yada}
\author{Masatoshi Sato}
\author{Yukio Tanaka}

\address{Department of Applied Physics, Nagoya University, Nagoya 464-8603, Japan}

\end{frontmatter}


\section{Introduction}
\label{intro}

Topological superconductors (TSCs) are a new state of matter which is characterized by the nonzero topological numbers of the bulk wave functions \cite{tanaka12jpsj, qi11rmp, wilczek09}. From the bulk-boundary correspondence, there exist topologically protected gapless Andreev bound states (ABSs) in TSCs. In particular, the superconductivity infers that the gapless ABSs are their own antiparticles, thus Majorana fermions. The realization of Majorana fermions in condensed matter physics is of particular interest because of their novelty as well as the possible application for quantum computing. 
The recent discovered superconductor Cu$_x$Bi$_2$Se$_3$ \cite{hor10} is a candidate of the TSC. The parent material Bi$_2$Se$_3$ is a topological insulator with a topologically protected gapless Dirac fermion on its surface. With doping Cu, however, the superconductivity appears. From the Fermi surface structure of the material, it was predicted that the superconducting topological insulator (STI) Cu$_x$Bi$_2$Se$_3$ must be a TSC \cite{fu10} if time-reversal invariant odd-parity superconductivity is realized \cite{sato09, sato10}. Possible ABSs specific to this material have been predicted \cite{hao11, hsieh12, yamakage12}. Indeed, point-contact spectroscopy experiments \cite{sasaki11,kirzhner12} have reported a zero-bias conduction peak (ZBCP) which signifies gapless ABSs \cite{hu94, tanaka95, kashiwaya00}. 
In addition, the superconducting proximity effect on topological insulators has been studied \cite{koren11, koren12, yang12, zareapour12}.
They also show a ZBCP.

The purpose of this work is to present a theory of the tunneling conductance for the STI. For one of the possible pairings, the ABS has a linear energy dispersion. In this case, surface density of states (SDOS) does not always express the tunneling conductance in normal metal (N)/superconductor junction \cite{yamashiro97, honerkamp98, matsumoto99, iniotakis07, eschrig12}, and the resulting tunneling conductance depends on the transparency at the interface. Moreover, the Balian-Werthamer (BW) state \cite{balian63}, which is a three-dimensional full-gap TSC and supports dispersive gapless ABSs, never shows a ZBCP \cite{asano03}. The relation between SDOS in three-dimensional full-gap superconductors and ZBCPs is non-trivial. Thus, in order to pursue the origin of the ZBCP in the STI, one needs to calculate directly the tunneling conductance of the N/STI junction.
The tunneling conductance for two of the possible pairings has been already studied in Ref. \cite{yamakage12}.
In this work, we calculate it for all the possible pairings. 
Thus, one can directly compare the theoretical features with the experimental ones. 
This is useful for determination of the pairing symmetry of Cu$_x$Bi$_2$Se$_3$. 
Furthermore, the STI with dispersive gapless ABSs can show a ZBCP, due to the enhancement of the SDOS induced by the surface-state transition in the momentum space. 
Especially, we show that energy of the ABS is proportional to the cube of the momentum at the transition point, where the corresponding SDOS diverges.
This is why a ZBCP appears in the TSC.
Note that the surface-state transition occurs since the parent material (a topological insulator) has a surface Dirac fermion and the mirror symmetry.
We explain evolution of energy spectrum of the ABS, focusing on the surface Dirac fermions and its mirror symmetry.

\section{Model}

A model Hamiltonian of a superconducting topological insulator is given by
\begin{align}
H_0(\bm k) &= 
-\mu +
m(\bm k) \sigma_x + v_z k_z \sigma_y + v \sigma_z (k_x s_y - k_y s_x),
\\
 H(\bm k) &=
H_0(\bm k) \tau_z
+ \hat \Delta_i \tau_x,
\\
m(\bm k) &= m_0 + m_1 k_z^2 + m_2 \left( k_x^2+k_y^2 \right),
\end{align}
where the basis is set to be $(c_{\sigma \uparrow}(\bm k),  c_{\sigma \downarrow} (\bm k),  -c^\dag_{\sigma,\downarrow}(-\bm k),  c^\dag_{\sigma \uparrow}(-\bm k))^{\rm T}$, $\sigma = \pm$ and $s=\uparrow, \downarrow$ denote the orbital and spin, $\sigma_i$, $s_i$, and $\tau_i$ are Pauli matrices in the orbital, spin, and Nambu spaces, respectively.
This system can have four types of momentum-independent pairing potential $\hat \Delta_i$ \cite{fu10}, which are summarized in Table \ref{pair}.
\begin{table}
\centering
\begin{tabular}{lccl}
\hline\hline
 Pairing potential $\hat \Delta_i$ & Irr. rep. & Parity & Energy gap
 \\
 \hline
 $\hat \Delta_{1 \mathrm a} = \Delta$ & \multirow{2}{*}{$A_{1g}$} & \multirow{2}{*}{even} & Isotropic full-gap
 \\
 $\hat \Delta_{1 \mathrm b} = \Delta \sigma_x$ & & & Anisotropic full-gap
 \\
 \hline
 $\hat{\Delta}_2 = \Delta \sigma_y s_z$ & $A_{1u}$ & odd & Anisotropic full-gap
 \\
 \hline
 $\hat \Delta_3 = \Delta \sigma_z$ & $A_{2u}$ & odd & Point nodes on $k_z$--axis
 \\
 \hline
 $\hat \Delta_4 = \Delta \sigma_y s_x$ ($\Delta \sigma_y s_y$)
 & $E_u$ & odd & Point nodes on $k_y$ ($k_x$) --axis.
 \\
 \hline\hline
\end{tabular}
\caption{Momentum-independent paring potentials $\hat \Delta_i$ in STIs with $\rm D_{3d}$ point group symmetry \cite{fu10}. The irreducible representation, parity eigenvalue, and structure of the energy gap are indicated.}
\label{pair}
\end{table}
%
%
%

The energy gap for each pair potential is shown in Fig. 1. 
\begin{figure}
\centering
\includegraphics{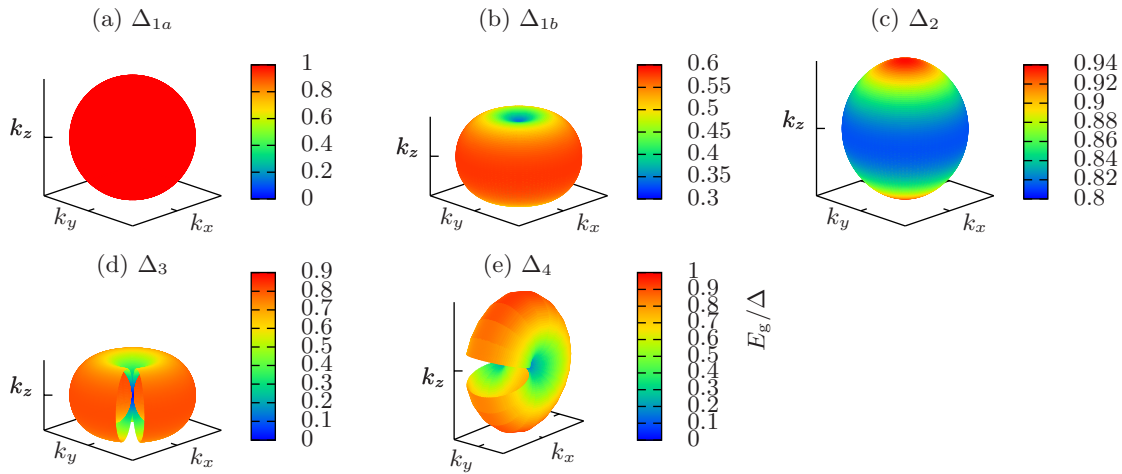}
\caption{Polar plots of energy gap $E_{\rm g}$ of the STIs. It is not plotted in a certain region for the cases (d) and (e), for visibility. The parameters are set to those of Bi$_2$Se$_3$, as follows. 
$m_0 = -0.28$eV, $m_1=20.18$eV\AA$^2$, $m_2=56.6$eV\AA$^2$, $v_z=3.09$eV\AA, $v=4.1$eV\AA, $\mu=0.5$eV, $\Delta=0.6$meV.}
\label{gap}
\end{figure}
For $\hat \Delta_{1 \rm a}$, an isotropic full-gap superconductor is realized. 
On the other hand, for $\hat \Delta_{1 \rm b}$ and $\hat \Delta_2$, the energy gap fully opens but is anisotropic. 
For $\hat \Delta_3$ and $\hat \Delta_4$, point nodes appear on the pole ($k_z$-axis) and on the equator ($k_x$ or $k_y$ --axis), respectively.

\section{Surface-state transition in the momentum space}

\begin{figure}
\centering
\includegraphics{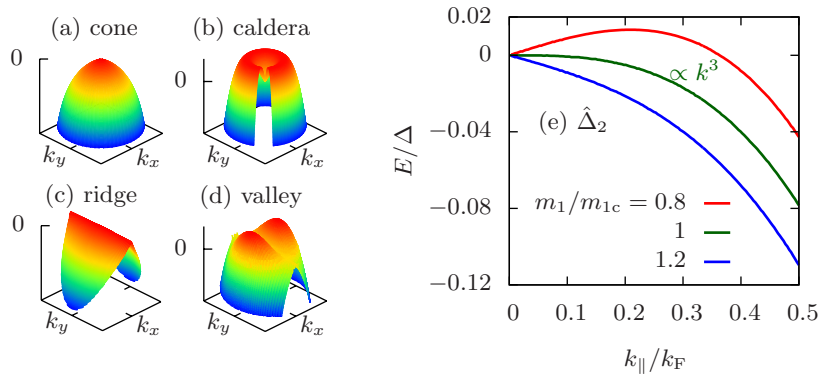}
\caption{
Energy spectra of the gapless ABSs. 
Only the half is shown.
The whole energy spectrum is given by mirroring with respect to $E=0$, preserving particle-hole symmetry. 
Pairing potentials are set to $\hat \Delta_2$ for (a) and (b), and to $\hat \Delta_4$ for (c) and (d). 
For (a) and (c) [(b) and (d)] we choose $m_1=20.18$eV\AA$^2$ ($m_1=5.65$eV\AA$^2$), where cone and ridge (caldera and valley) type energy spectra are realized, respectively.
(e): The energy spectra as a function of $k_\parallel/k_{\rm F}$ with $k_\parallel = (k_x^2 + k_y^2)^{1/2}$ for $\hat \Delta_2$ near the transition point ($m_1 \sim m_{1 \rm c} \equiv m_0 v_z^2/\mu^2$). $k_{\rm F}$ denotes the Fermi momentum for $k_z=0$.
}
\label{surface}
\end{figure}

 On the surface perpendicular to the $z$-axis, STIs with $\hat \Delta_2$ and with $\hat \Delta_4$ support gapless ABSs \cite{fu10,hao11,hsieh12,yamakage12}. 
Figure \ref{surface} shows energy spectra of the gapless ABSs of STIs with $\hat \Delta_2$ and $\hat \Delta_4$ for different values of $m_1$.
There are four types of ABSs, i.e., 
(a):
cone type for $\hat \Delta_2$, larger $\mu$ and $m_1$, whose energy spectrum is monotonic function of $k$, (b): caldera type for $\hat \Delta_2$, smaller $\mu$ and $m_1$, where the energy spectrum has zeros both at $k=0$ and at finite $k$, (c): ridge type for $\hat \Delta_4$, larger $\mu$ and $m_1$, where zero energy states appear between the two point nodes (on the $k_y$-axis), (d): valley type for $\hat \Delta_4$, smaller $\mu$ and $m_1$, where the energy remains zero on the $k_y$-axis but becomes a monotonic function of $k_x$ and zero-energy states appear at a finite momentum.

Differently from the other topological superconductors, these STIs can have twisted shape of energy spectra of gapless ABSs \cite{hao11,hsieh12,yamakage12} for smaller $\mu$ and $m_1$, as shown in Figs. \ref{surface}(b) and \ref{surface}(d). 
Zero-energy states at finite momenta stem from surface Dirac fermions in the normal state.
In other words, pairing potentials of $\hat \Delta_2$ and $\hat \Delta_4$ cannot open superconducting gap for the surface Dirac fermions, protected by the mirror symmetry \cite{hsieh12,yamakage12}, as explained below.
On the other hand, for much larger $\mu$ and $m_1$ where the Fermi level is much higher than the energy of surface Dirac fermions, the resulting energy spectrum of the gapless ABSs are not twisted since surface Dirac fermions are irrelevant in this energy regime.
Consequently, the transition from the twisted one to the conventional one occurs at the intermediate values of $\mu$ and $m_1$ as $m_{1} \mu^2 = m_0 v_z^2$ \cite{yamakage12}. 
Then, at this transition point, the energy spectrum of the ABS is proportional to $k^3$ [Fig. \ref{surface}(e)] and the corresponding SDOS diverges. 
In the vicinity of the transition point, the tunneling conductance yields a ZBCP due to the divergent SDOS at the transition point, as we will see in the next section.

Here, we explain  symmetry-protection of the zero-energy ABS  at a finite momentum for $\hat \Delta_2$, in terms of the mirror symmetry \cite{hsieh12, yamakage12}.
The system  has mirror symmetry as $s_x \tau_z H_0(-k_x,k_y,k_z) \tau_z s_x = H_0(k_x,k_y,k_z)$.
Then, $s_x \tau_z = \pm 1$ becomes a good quantum number for $k_x=0$.
The particle [$H_0(\bm k)$] and hole [$-H_0(\bm k)$] components of a surface Dirac fermion for $\Delta=0$ have the different mirror eigenvalues.
On the other hand, $\hat \Delta_2 = \Delta \sigma_y s_z$ is even under the mirror reflection as $s_x \tau_z \hat \Delta_2 \tau_z s_x = \hat \Delta_2$.
As a result, due to mismatch of the mirror eigenvalues, $\hat \Delta_2$ cannot open the superconducting gap of the surface Dirac fermions.
Namely, when the system has surface Dirac fermions at the Fermi level, they still remain gapless and turns into zero-energy ABSs.
In this picture, the zero-energy ABSs are located at the Fermi momentum of the surface Dirac fermion $k_{\rm DF}$ in the normal state. 
As one increases $\mu$ and $m_1$, $k_{\rm DF}$ decreases.
For certain values of $\mu$ and $m_1$ with $k_{\rm DF} =k_{\rm F}$, the surface Dirac fermions are gone from the Fermi level.
However, the ABSs remains gapless since the ABSs with the different mirror eigenvalues are not hybridized.
Further increasing $\mu$ and $m_1$, the momentum of the zero-energy ABSs decreases to 0, then the shape of the energy spectrum changes from twisted to non-twisted, i.e., the surface-state transition occurs.
Note that the zero-energy ABSs protected by the mirror symmetry on the $k_y$-axis are robust even if the higher order terms of $k$, e.g., the warping term \cite{fu09}, is taken into account.

\section{Tunneling conductance}
Now, we discuss the tunneling conductance of  N/STI junctions. 
Hamiltonian of the normal metal is set to 
$H_{\rm N}(\bm k) = (k_x^2+k_y^2+k_z^2)/(2m_{\rm N}) - \mu_{\rm N}$. 
We calculate the tunneling conductance by using BTK formula \cite{yamakage12} for various $\mu_{\rm N}$, 
which  
corresponds to transmissivity of the junction.
When $\mu_{\rm N}/{\mu} \sim 0.6$, the transmissivity of the junction is high $G_{\rm N}/G_0 \sim 1$ since the magnitudes of the Fermi momenta of the normal metal and of the STI are the same. 
Here, conductance of a perfect transmission is given by $G_0 = k_{\rm N F}^2 L^2 /(4 \pi) \times e^2/h$, where $k_{\rm N F}$ is Fermi momentum of the normal metal and $L^2$ denotes cross section of the system.
As one increases $\mu_{\rm N}$, the magnitude of Fermi momentum of the normal metal increases and then the transmissivity decreases.

\begin{figure}
\centering
\includegraphics{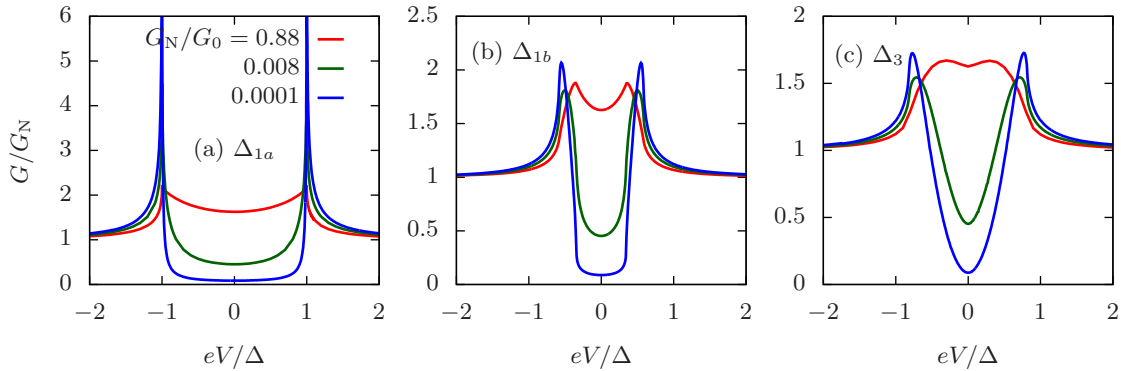}
\caption{Tunneling conductance as a function of bias voltage $V$.
$G_{\rm N}/G_0$ denotes the normalized conductance of the normal state ($\Delta=0$), which corresponds to the transmissivity.
 The parameters of STI are the same as in Fig. \ref{surface}, and $\mu_{\rm N}/\mu = 0.6, 60$, and 1200 for $G_{\rm N}/G_0 = 0.88, 0.008$, and 0.0001, respectively.
Effective mass of electron in the normal metal is taken to be $m_{\rm N} = 1/m_2$.
}
\label{conductance1}
\end{figure}

\begin{figure}
\centering
\includegraphics{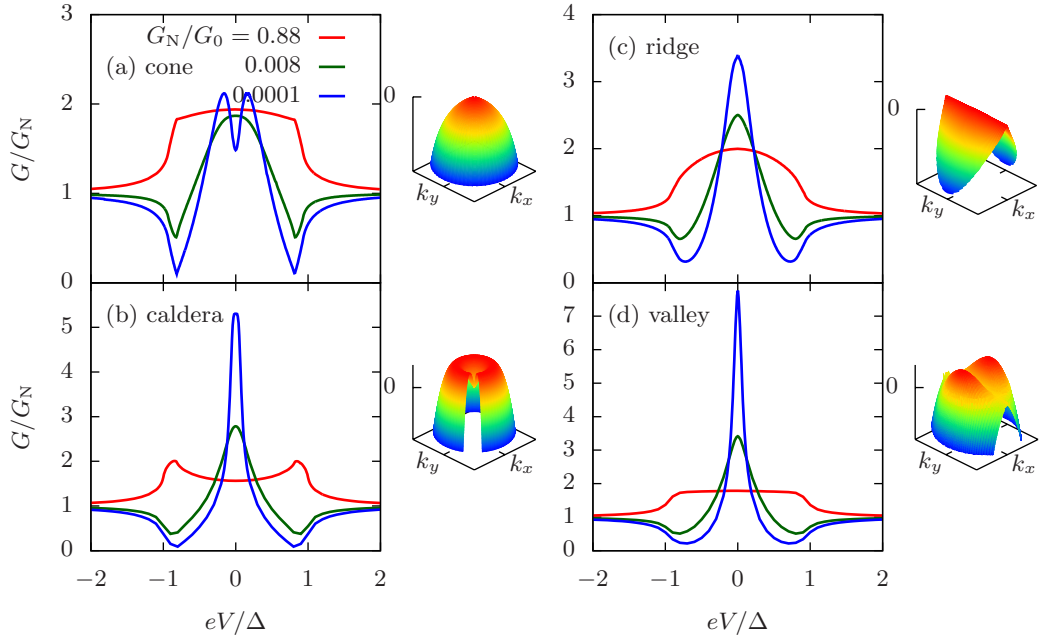}
\caption{Tunneling conductance and the corresponding gapless ABSs of the STIs.
The parameter are the same as in Figs. \ref{surface} and \ref{conductance1}.
}
\label{conductance2}
\end{figure}

Figures \ref{conductance1} and \ref{conductance2} show the tunneling conductance $G$ normalized by that of the normal ($\Delta=0$) state $G_{\rm N}$.
The line shapes of the tunneling conductances for $\hat \Delta_{1 \rm a}$ and $\hat \Delta_{1\rm b}$ are the same as that of the conventional s-wave superconductor, as shown in Figs. \ref{conductance1}(a) and \ref{conductance1}(b). 
For $\hat \Delta_3$, the system has point nodes on the $k_z$-axis [Fig. \ref{gap}(d)] and has no gapless ABS on the surface perpendicular to $z$-axis. 
The corresponding tunneling conductance is given by $G \propto V^2$  with  $V$ being bias voltage [Fig. \ref{conductance1}(c)].

On the other hand,
for $\hat \Delta_2$ with a conventional cone type ABS [Fig. \ref{conductance2}(a)], a ZBCP is realized for the intermediate transmissivity ($G_{\rm N}/G_0 = 0.008$). 
It is noted that a ZBCP never appears for the BW state \cite{asano03}, which is a well-known three-dimensional TSC. In the STI, as discussed in the previous section, the SDOS diverges at the zero-energy for the surface-state transition point. Therefore, in the vicinity of this transition point, the magnitude of SDOS near the zero-energy becomes large. 
This results in a ZBCP in the intermediate transparent case. 
In the low transparent limit ($G_{\rm N}/G_0 = 0.0001$), the line shape of the tunneling conductance completely converges to that of the SDOS. Thus the line shape of tunneling conductance for $G_{\rm N}/G_0 = 0.0001$ forms a double-peak and a zero-bias dip [Fig. \ref{conductance2}(a)], which is the same as that of the BW state. 
%
For  $\hat \Delta_2$ with the caldera-type ABS,
the corresponding SDOS near the zero energy becomes larger hence the height of the resulting ZBCP becomes larger [Fig. \ref{conductance2}(b)]. In addition, the ZBCP survives in a lower transmissivity ($G_{\rm N}/G_0 = 0.0001$).

 Finally, for $\hat \Delta_4$, 
the resulting tunneling conductance shows a ZBCP for any transmissivity [Figs. \ref{conductance2}(c) and \ref{conductance2}(d)], 
since zero-energy ABSs appear between the two point nodes, i.e., a flat band on the $k_y$-axis appears.
For smaller $m_1$ and $\mu$, the gapless ABSs changes from the ridge to caldera types. 
Similarly to that of caldera type ABS of $\hat \Delta_2$ [Fig. \ref{conductance2}(b)], the resulting tunneling conductance becomes prominent [Fig. \ref{conductance2}(d)].

\section{Summary}

In this work, we have clarified tunneling conductance of normal metal / STI junctions for possible pair potentials. 
Only STIs with $\hat \Delta_2$ in the intermediate transmissivity and $\hat \Delta_4$ show a ZBCP. 
The presence of ZBCP is possible in the STI with $\hat \Delta_2$ although the energy spectrum of gapless ABS is dispersive. 
A ZBCP in a STI with $\hat \Delta_2$ originates from the surface-state transition, which comes from surface Dirac fermions in the normal state, i.e., topological insulator.
This is not the case in the BW state. 
Our obtained results serve as a guide to explore three-dimensional topological superconductor hosting Majorana fermions.

\section*{Acknowledgments}
This work is supported by the ``Topological Quantum Phenomena" (No. 22103005) Grant-in Aid for Scientific Research on Innovative Areas from the Ministry of Education, Culture, Sports, Science and Technology (MEXT) of Japan.





\bibliographystyle{model1a-num-names}
\bibliography{yamakage}







\end{document}